\documentclass{taira}

\usepackage{amssymb}
\usepackage{amsmath}
\usepackage{graphicx}
\usepackage{lipsum}
\usepackage{tabu}
\usepackage{multirow}
\usepackage{bm}
\usepackage{mathrsfs}
\begin{document}

\shorttitle{Acoustic response of turbulent cavity flow} 
\shortauthor{Q. Liu \& D. Gaitonde} 

\title{Acoustic response of turbulent cavity flow using resolvent analysis} 

\author{Qiong Liu
\corresp{\email{liu.9292@osu.edu}} and
Datta Gaitonde
}

\affiliation{
Department of Mechanical and Aerospace Engineering, The Ohio State University, Columbus \\
OH 43220, USA 

}

\maketitle

\begin{abstract}
Fluid-acoustic interactions are important in a variety of applications, and typically result in adverse effects.
We analyze the influence of Mach number on such interactions and their input-output characteristics by combining resolvent analysis with Doak's momentum potential theory.
The specific problem selected is the flow over an open cavity of $L/D=6$ at $Re=10,000$ and $M_\infty=0.6$ and $1.4$, respectively. 
The resolvent forcing and response modes of the time- and spanwise-averaged Large Eddy Simulations are first obtained at each Mach number.
The response modes are then decomposed into their hydrodynamic, acoustic and thermal components.
Although the results depend quantitatively on Mach number, some trends remain consistent.
In particular, at lower frequencies, the acoustic component appears primarily at the trailing edge of the cavity. 
When the frequency is increased, the primary acoustic response moves towards the leading edge and overlaps with its hydrodynamic component, indicating greater influence on the flowfield.
Inspired by actual cavity flow control, the forcing is then localized to two regions -- the leading-edge and front wall of the cavity -- and also filtered to consider notional actuators that can separately introduce each component of velocity, density, and temperature forcing, respectively. 
Among these different types of actuation perturbations, regardless of Mach number, streamwise velocity forcing achieves the largest energy amplification when placed at the leading edge, with considerable reduction in effectiveness when placed on the front wall. 
A suitable ratio is defined to assess the relative acoustic versus hydrodynamic modification.
The frequency where this ratio is largest is often slightly higher than that associated with the energy amplification peak. 
For both subsonic and supersonic cavity flows, beyond a certain forcing frequency threshold value, the nature of the acoustic versus hydrodynamic response becomes independent of the forcing type; however, the amplification continues to be strongly impacted by the forcing frequency.
This reinforces the importance of the frequency for energy amplification mechanisms. 
Overall, this work provides an alternative approach to examine input-output flow-acoustic characteristics where resonance is dominant and provides a means to evaluate the relative effectiveness of different types and locations of actuation.
\end{abstract}

\maketitle

\section{Introduction}
The wide range of applications where cavity flows are important has motivated numerous efforts to understand their physics and to design effective flow controls.
These inquiries started from 1950's~\citep{krishnamurty1955acoustic} with both experimental~\citep{roshko1955some,Rossiter1964,heller1971flow,ukeiley2004suppression,murray2009properties,wagner2015fluid} and complementary numerical simulations~\citep{plumblee1962theoretical,shaw1979suppression,larcheveque2007large,liu2016linear}. 
A dominant mechanism of interest is fluid-acoustic resonance~\citep{Powell1953,Powell1961,Rossiter1964}, which drives self-sustained oscillations in compressible cavity flows. 
As disturbances grow from the cavity leading edge, vortical structures are formed over the shear-layer and impinge on the cavity trailing edge, scattering acoustic waves. 
These disturbances then propagate upstream and interfere with the unsteadiness of vortical structures. 
Such interactions between flow and acoustic waves produce a feedback loop type of oscillation~\citep{rowley2006dynamics,cattafesta2008active,lawson2011review}. 
The resonant acoustic mode may enhance or suppress the flow-acoustic resonance, which results in mode selection or mode switching over the cavity~\citep{rowley2006dynamics}. 
Moreover, a centrifugal type instability closely related to the recirculation inside the cavity can coexist with the resonant dynamics. 
As the cavity becomes three-dimensional, the centrifugal instability promotes fluid exchange inside the cavity~\citep{faure2007visualizations,bres2008three,de2014three,sun2017biglobal,picella2018successive}.

Usually, fluid-acoustic resonance leads to significant sound pressure levels, which adversely affect the performance and threaten structural safety. 
There is therefore strong interest in developing flow control techniques with the objective of attenuating pressure fluctuations over the cavity by suppressing acoustic tones~\citep{cattafesta2008active,rowley2006dynamics}. 
Additionally, the cavity control problem serves as a model problem that is frequently used in multidisciplinary flow control research, and as an exemplar of flow-acoustic resonance in different flows of engineering interest.

Passive and active control approaches have been successfully implemented in cavity flows. 
The former requires changes to the configuration, while the latter introduces external energy into the system with actuators~\citep{cattafesta1997active,cattafesta2008active,cattafesta2011actuators}.
Passive control is generally cheaper and simpler; examples include trailing edge modifications\citep{Pereira1994influence,Zhang1998effect}, leading-edge fences and spoilers~\citep{heller1971flow,shaw1979suppression,ukeiley2004suppression}.  
However, it can be limiting at operating conditions where control is not required. 
In contrast,  active flow control techniques can adapt to flow conditions, and are further sub-divided into open and close loop variants~\citep{barbagallo2009closed,Brunton2015closed}. 
Closed-loop control implies the active exploitation of a feedback loop, in which the actuation is modified by sensor signals~\citep{distefano2012feedback}, while open-loop control does not adapt to the flow response.

Active open-loop controls are also common in flow control studies, and different actuator types have been explored.
\citet{Williams2000closed} performed an experimental control study using a zero-net-mass type of actuation at the leading edge of the cavity. 
The experiment was conducted using forcing jet directions of $0^\circ$, $45^\circ$, and $90^\circ$, respectively, to the free-stream flow. 
The control result showed that forcing in-line with the free-stream was more effective at suppressing resonance tones, than the other directions.  
\citet{ukeiley2007control} sought the optimal angle for leading-edge blowing in a supersonic ($M_\infty=1.5$) three-dimensional cavity flow with injection actuation at the leading-edge of the cavity. 
The study compared the control effects from three injection directions of $0^\circ$, $45^\circ$, and $90^\circ$ with the reference direction of free-stream flow. 
The most effective control case was the mass injection actuation with a blowing direction of $90^\circ$. 
Other types of actuators employed include those using plasma effects~\citep{chan2007attenuation}; the use of such actuators at the leading edge of the cavity was observed to be effective for acoustic noise reduction. 
Thus, the different control approaches require separate assessments, since effectiveness depends on the specifics of the control perturbation.

Also important for control effectiveness is the location where actuators are placed relative to the location of instability growth and natural feedback.
Typically, to foster stronger interactions with the shear layer, actuators are placed at the leading edge or front wall of the cavity.
\citet{lamp1997computation} simulated the control effect on a supersonic cavity flow using a blowing jet on the front wall at a fixed angle of $45^\circ$ to the free-stream. 
They found that the effectiveness of the suppression depended strongly on the amplitude and frequency and only weakly on the phase angle and duty cycle of the actuation. 
Front-wall blowing jet did not affect the timing of events within the cavity; however, controlled results using leading edge mass blowing~\citep{ukeiley2007control} reduced the extent and magnitude of the reverse flow region inside the cavity.
The placement of control actuation thus clearly has a strong impact on the affected regime.

Although these control studies provide significant information about the sensitivity of cavity flow, several interesting questions arise which motivate the present study,
\begin{enumerate}
\item What is the effect of choosing different types of forcing? Specifically, how do mass and momentum perturbations through velocity forcing differ from those arising due to thermal or pressure forcing in terms of the flow-acoustic interaction?
\item How can the influence of actuator location, whether at the leading edge or front wall, be characterized?
\item How are the results affected by flow parameters, specifically the Mach number?
\end{enumerate}

We consider two sequential components to answer these questions in the context of the flow past a cavity.
The first evaluates input-output characteristics of the flow taking recourse to resolvent analysis as a modal approach.
Recent work \citep{liu2020unsteady} shows promising control results by adopting resolvent analysis~\citep{jovanovic2005componentwise,mckeon2010critical,schmid2012stability,gomez2016reduced,yeh2018resolvent}. 
Specifically, a zero-net-mass actuation design, applied at the leading edge, with prescribed frequency and spanwise spacings were analyzed with the resolvent approach.
The control achieved more than $50\%$ reduction in pressure root-mean-square (r.m.s.) values  and also significantly suppressed the unsteadiness content over the cavity. 
The control mechanism was traced to the generation of streamwise vortices near the 3D actuation, which prevents the formation of large spanwise coherent structures; this subsequently mitigates the adverse effects of trailing edge impingement. 

Since the acoustic wave plays an indispensable role in the self-sustained resonance mechanism, it becomes important to understand its influence on the  input-output characteristics of the flow. 
To properly consider the flow-acoustic interaction therefore, it is advantageous to analyze the dynamics of a suitably defined purely acoustic component of fluctuations, if such a separation can be achieved.
This naturally leads to consideration of a Kovasznay-type of fluctuation splitting~\cite{kovasznay1953turbulence} into three components: hydrodynamic, acoustic, and thermal, associated with fluctuations of vorticity, pressure and entropy, respectively.
The difficulties of performing such a splitting with the Kovasznay approach when the mean state is non-uniform or the dynamics are non-linear may be discerned from \citet{chu1958non}.
A major reason for the difficulty stems from the association of independent scalar variables with each type of component.

A more elegant method based on Doak’s momentum potential theory (MPT)~\citep{doak1989momentum} is adopted in this work.
The method exactly splits a single variable ($\rho \boldsymbol{u}$) into its acoustic (irrotational-isentropic), hydrodynamic or vortical (rotational) and a thermal or entropic (irrotational-isobaric) component. 
To distinguish this decomposition from the Kovasznay approach, we designate these as fluid-thermodynamic (FT) components.
The splitting is exact regardless of non-linearities or variations in the underlying mean flow and has been successfully applied to model problems~\citep{daviller2009flow,jordan2013doak}, jets and transition~\citep{unnikrishnan2016acoustic,unnikrishnan2019interactions}.

The combination of the resolvent framework with Doak's decomposition is thus a very effective strategy to further extract insights into the dynamics of actuation response. 
Specifically, we examine the input-output characteristics from resolvent-informed forcing at different frequency and wavenumber combinations in the context of their hydrodynamic, acoustic, and thermal content.
This then facilitates specific conclusions on the behavior of the acoustic mode and its interactions with the other components that result in the observed near-field sound signature and the features inside the cavity and the end wall.
The use of MPT with input-output analysis to provide insightful information on the fluid-thermodynamic response to forcing has been illustrated by ~\citet{houston2020aero}, who  separated disturbances in a hypersonic boundary layer over blunt cone into their vortical, thermal and acoustic components. 
A key finding was that the vortical output is more sensitive to the wall-normal forcing. 

The results from the underlying flow-acoustic interaction feed naturally into the answer to the second, more practical, question above regarding actuator placement.
Usually, the forcing location is chosen to be close to the leading edge, which is considered to be the most sensitive region to alter the shear layer. 
Consequently, we restrict the input forcing to two distinct locations: leading-edge and front wall. 
The other important property concerns the nature of the perturbation introduced (mass or energy injection for example), which varies with the actuator employed.  
To isolate the influence of this variable, we consider idealized inputs corresponding to different types of perturbations, one at a time.
Thus, the independent effects of velocity components, density and temperature perturbations are evaluated on the overall desired flow-acoustic interaction.

Finally, the response and effectiveness of perturbations depends substantially on the flow parameters, particularly Mach number.
Control of supersonic ($M_\infty=1.44$) and subsonic ($M_\infty=0.6$) cavity flow were experimentally conducted by Lusk et al~\cite{lusk2012leading} and Zhang et al~\cite{zhang2019suppression}, respectively. 
The results showed that the most effective control actuation for the subsonic flow with the slot length was three times longer than that for supersonic control.
The Mach number is thus clearly influential, not only because of the change in the nature of resonance and acoustic emission, but also because of the differences in response to control perturbations.  
To address this aspect, in this work, we analyze the results for two Mach numbers, one subsonic ($M_\infty=0.6$) and the other supersonic ($M_\infty=1.4$).

In summary, the present study extracts the flow-acoustic response of supersonic and subsonic cavity flows by identifying input-output characteristics in the context of their fluid-thermodynamic content to provide physical insights for flow control design. 
The details of the numerical setup and methodology are documented in section~\ref{sec:approach}. Inspired by previous control studies, we apply resolvent analysis on time-averaged base flows and process the results using Doak's decomposition.
The focus of section~\ref{sec:results} is on the variation of energy amplification and its flow-acoustic response due to frequency, wave number, different types of global/localized forcings and Mach number. 
Concluding remarks are offered in section~\ref{sec:conclusion}. 

\section{Problem description and approach}
\label{sec:approach}
\subsection{Problem setup}

\begin{figure*}
\centering
\includegraphics [width=\textwidth,trim={0 0cm 0 0cm},clip]{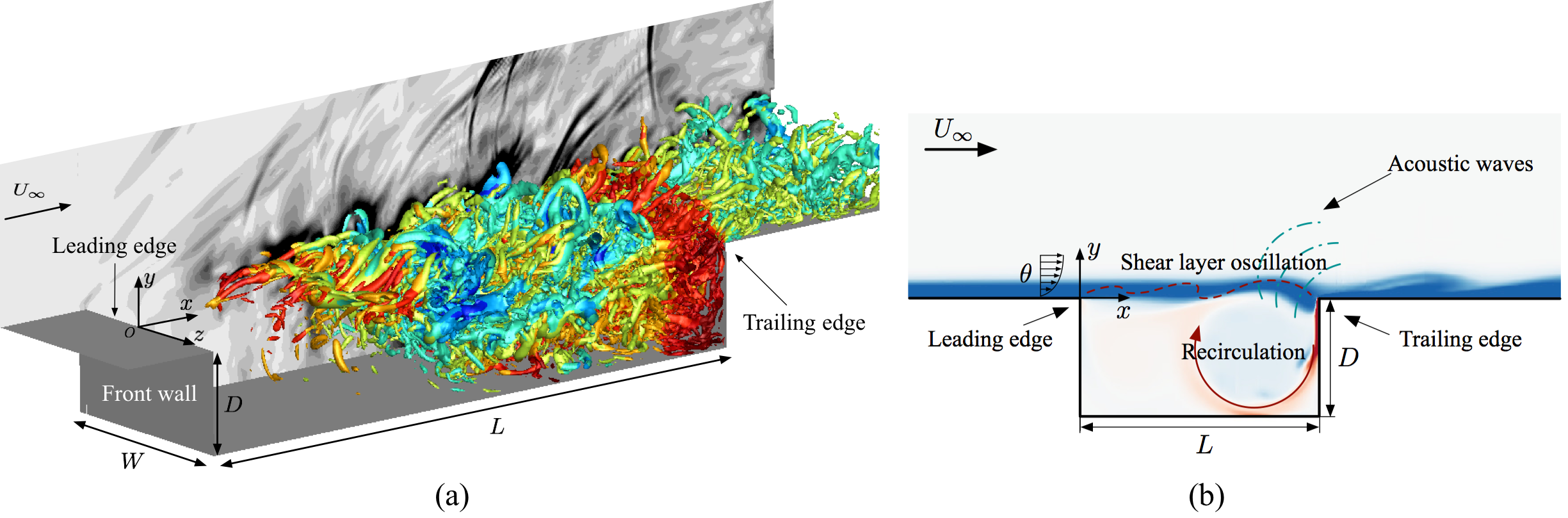}
\caption{Problem description: (a) compressible flow over a rectangular cavity with $L/D = 6$ and $W/D = 2$ at $Re=10,000$. The computational setup is shown along with the instantaneous $Q-\text{criterion}=10$ colored by pressure fluctuation at $M_\infty= 1.4$. The density gradient magnitude is shown in gray color. (b) Schematic of flow-acoustic feedback loop in compressible cavity flow (not to scale).}
\label{fig:Geo}	
\end{figure*}

We consider the turbulent flow over a rectangular cavity at a subsonic ($M_\infty \equiv u_\infty/a_\infty=0.6$) and a supersonic ($1.4$) Mach number, where $a_\infty$ is the sound speed and $u_\infty$ is the free-stream velocity.
The cavity length-to-depth ratio is $L/D = 6$. 
FIG.~\ref{fig:Geo} shows an instantaneous snapshot of the compressible flow over the cavity as well as a schematic of flow-acoustic feedback loop in compressible cavity flow.
The Reynolds number is $Re \equiv \rho_\infty u_\infty D/\mu_\infty=10,000$, where $\mu_\infty$ and $\rho_\infty$ are the free-stream dynamic viscosity and density, respectively. All variables are non-dimensionalized, lengths by the cavity depth $D$, temperature by $T_\infty$, pressure by $\frac{1}{2} \rho_\infty u_\infty^2$, density by $\rho_\infty$, and time by $D/u_\infty$. 

The numerical setup~\citep{sun2018effects,liu2020unsteady,sun2019resolvent} has been validated with the companion experimental study~\citep{zhang2019suppression}.
The basic state used in the resolvent analysis is the time- and spanwise-averaged flow obtained from an LES performed by the compressible flow solver \emph{CharLES} \citep{Khalighi:ASME2011,Bres:AIAAJ17}. 
The solver is based on a second-order finite-volume discretization and a third-order Runge--Kutta time integration scheme. 
In the supersonic case, the Harten-Lax-van Leer contact scheme \citep{Toro:94} is used to capture shocks. 
A refined mesh of $488\times200\times128$ points is used in the $x$, $y$, and $z$ directions around the cavity region $(x,y,z)/D\in [-1, 7]\times[-1,1]\times[-1,1]$. 
The origin of coordinate is set at middle of the leading edge.

The turbulent boundary layer thickness upstream of the cavity lip is specified to be $\delta_0/D=0.167$ based on the companion experiments~\cite{george2015control,zhang2019suppression}.
A one-seventh power law velocity profile is imposed, with random superimposed Fourier modes to simulate unsteady fluctuations entering the cavity interaction \citep{bechara1994stochastic}. No-slip and adiabatic wall boundary conditions are specified along the cavity walls. 
Spanwise periodic boundary conditions are enforced. 
A sponge boundary condition is applied at the farfield boundary, while waves exiting the outlet boundaries are damped to prevent numerical reflections. 

\subsection{Resolvent analysis}\label{sec:Resolvent_methodology}
The input-output characteristics of the turbulent cavity flows at each Mach number are first discussed, as obtained from a resolvent analysis. 
For this, the flow variables are decomposed into the spanwise- and time-averaged base state $\bar{\bm{q}}(x,y)\equiv[\bar{\rho},\bar{u},\bar{v},\bar{w},\bar{T}]$  and statistically stationary fluctuating components $\bm{q}'(x,y,z,t)\equiv[\rho',u',v',w',T']$. 
The resulting fluctuation Navier-Stokes equation can be expressed as an input-output system~\citep{jovanovic2005componentwise,mckeon2010critical,schmid2012stability},
\begin{equation}
\label{eqn:decom_NS}
    \frac{\partial \bm q'}{\partial t}=\mathcal{L}(\bar{\bm q})\bm q'+\mathcal{M}\bm f'
\end{equation}
where $\mathcal{L}(\bar{q})$ is the Navier-Stokes operator linearized about the base state $\bar{\bm{q}}$. 
The finite-amplitude nonlinear terms are incorporated in ${\bm f}'$. 
$\mathcal{M}$ is the coupling matrix as further discussed below.

The spanwise-homogenous property of the base state enables the use of a modal ansatz with spanwise wavenumber $\beta$ and temporal frequency $\omega$:

\begin{equation}\label{eqn: ansatz}
{\bm q}'(x,y,z,t)=\hat{\bm q}(x,y)e^{-\text{i}\omega t-\beta z},~ 
{\bm f}'(x,y,z,t)=\hat{\bm f}(x,y)e^{-\text{i}\omega t-\beta z}
\end{equation}

Inserting these into equation (\ref{eqn:decom_NS}), we obtain     
\begin{equation}\label{eqn: Resolvent}
	\hat{{\bm q}} 
	=[-\text{i}\omega \mathcal{I} - \mathcal{L}({\bm{\bar q}},\beta)]^{-1}\mathcal{M}\hat{{\bm f}}
\end{equation}
where $\mathcal{R}=[\text{i}\omega \mathcal{I} - \mathcal{L}({\bm{\bar q}},\beta)]^{-1}\mathcal{M}$ is the resolvent operator, which serves as a transfer function between the input $\hat{\bm f}$ and the corresponding output $\hat{\bm q}$ for a given flow state ($\bm{\bar{q}}$) and modal parameters ($\beta$ and $\omega$). 

The energy amplification of the system may be evaluated from the ratio of output to input energy $\frac{||\hat{\bm{q}}||_E}{||\hat{\bm{f}}||_E}$, where $||\cdot||_E$ is an energy norm. 
A singular value decomposition (SVD) of the resolvent operator facilitates a ranking of the energy amplification ratio in descending order. 
For compressible flow, the compressible energy norm~\citep{chu1965energy} is used:
\begin{displaymath}
E=\int_S \left[ 
	\frac{\bar{a}^2 \rho^2}{\gamma \bar{\rho}}+\bar{\rho}({u}^2+{v}^2+{w}^2) + \frac{\bar{\rho}C_v T^2}{\bar{T}}
	\right] {\text d}s
\end{displaymath}
where $S$ is the domain of interest.  
This yields
\begin{equation}\label{eqn: Resolvent_2}
	W^{\frac{1}{2}}\mathcal{R}W^{-\frac{1}{2}} = \mathcal{Q} \Sigma  \mathcal{F}^\ast,
\end{equation}
in terms of the weight matrix, $W$, based on the compressible energy norm $E$. 
The matrix $\mathcal{Q}=[\hat{\bm q}_1, \hat{\bm q}_2,\dots, \hat{\bm q}_n]$ holds the set of optimal response directions and $\mathcal{F}=[\hat{\bm f}_{1}, \hat{\bm f}_{2},\dots, \hat{\bm f}_{n}]$ contains the corresponding forcing directions, where $\hat{\bm q}_i=(\hat{\rho}_r,\hat{u}_r,\hat{v}_r,\hat{w}_r,\hat{T}_r,)$ and $\hat{\bm f}_i=(\hat{\rho}_f,\hat{u}_f,\hat{v}_f,\hat{w}_f,\hat{T}_f,)$ with $n$ is number of solved singular values.
The superscript $\ast$ denotes the Hermitian transpose. 
The singular values $\Sigma=\text{diag}(\sigma_1,\sigma_2,\dots,\sigma_n)$ represent the energy amplification (gain) between response and forcing modes.

For computational efficiency, the resolvent analysis is performed on the computational domain which has a downstream and farfield extent of $5D$ and grid size of $50,358$ cells. 
This is reasonable because the primary flow physics of interest occur over the shear layer region and inside the cavity. 
A smaller size computational domain and grid balances the computational efficiency and flow dynamics of the desired investigation. 
The SVD is performed using the ARPACK package with a Krylov space of $12$ vectors and a residual tolerance of $10^{-7}$. 
The results converge to at least $7$ significant digits and are verified to be $O(1\%)$ of accuracy with respect to the domain size and mesh resolution.

Since the base flow is unstable, it becomes crucial to highlight amplifications that occur on a shorter time scale than those associated with the asymptotic behavior observed with classical instability theory.
This consideration aids in achieving the main objective of finding preferred energy transfer mechanisms from the mean flow to the fluctuation field, which are necessary to provide physical insight into potential flow control strategies. 
To achieve this objective, the discounting technique ~\citep{jovanovic2004modeling,yeh2018resolvent} is employed to obtain forcing and response modes.
The method introduces a free parameter, denoted the discounting parameter $\kappa$, the choice of which is predicated on information about the most unstable growth rate as obtained from the stability analysis. 
Other, more physical techniques, to address the unstable linear operator may be found in~\citep{schmidt2017spectral,pickering2019eddy}.

\subsection{Doak’s momentum potential theory}
The distribution of the response modes into their fluid-thermodynamic content is performed using  Doak’s momentum potential theory~\citep{doak1989momentum} in frequency-wavenumber domain.
The corresponding time-domain implementation for fluctuations obtained from LES has been discussed~\cite{unnikrishnan2016acoustic,prasad2020study} for various free jet flows. 
The frequency-wavenumber domain  is more suitable for application to the response mode.
A crucial feature of the approach is to adopt a vector quantity, the momentum density, $\rho \bm{u}$, as the primary dependent field on which to perform the decomposition. 
This is expressed as a sum of solenoidal and irrotational components according to Helmholtz's theorem,
\begin{equation}\label{eqn: decomposition1}
{\rho \bm u}= {\bm B} - \frac{\partial {\psi}}{\partial \bm x},~~~\frac{\partial {\bm B}}{\partial \bm x}=0
 \end{equation}
where $\rho$ is the density and $\bm u$ is the velocity vector. 
The solenoidal and irrotational components are ${\bm B}$ and $- \partial {\psi}/\partial \bm x$, respectively. 

Upon substitution of equation~(\ref{eqn: decomposition1}) into the continuity equation, the following relationship is obtained
\begin{equation}\label{eqn: continuteyequation}
\frac{\partial {\rho}}{\partial t}-\frac{\partial^2 {\psi}}{\partial \bm x^2}=0.
\end{equation}
By Reynolds decomposing the flow variables into a time-averaged state and fluctuation components, the retained fluctuation density and scalar momentum potential gradient become
\begin{equation}\label{eqn: flucceqn}
\frac{\partial {\rho'}}{\partial t}=\frac{\partial^2 {\psi'}}{\partial \bm x^2}
\end{equation}
For the single-chemical-component flow, the density is considered as a function of pressure $p$ and entropy $S$. 
Hence, $\partial \rho'/\partial t$ can be splitted as the sum of $1/c^2\partial p'/\partial t$ and $\rho_s\partial S'/\partial t$, where $c$ is local sound speed. 
This effectively splits the irrotational field into its acoustic and thermal components.
Equation~(\ref{eqn: flucceqn})  then becomes
\begin{equation}\label{eqn: flucceqn2}
 \frac{\partial^2 {\psi'}}{\partial \bm x^2}=\frac{1}{c^2}\frac{\partial p'}{\partial t}+\rho_s\frac{\partial S'}{\partial t}.
\end{equation}
Thus $\psi'$ can be written as the sum of acoustic $\psi'_A$ and thermal $\psi'_T$ components, where 
$\partial^2\psi'_A/\partial {\bm x}^2=1/c^2\partial p'/\partial t$ and $\partial \psi'_T/\partial {\bm x}^2=\rho_s\partial S'/\partial t$.

By again considering the spanwise homogeneous nature of the flow problem, modal expressions~(\ref{eqn: ansatz}) with spanwise wavenumber $\beta$ and temporal frequency $\omega$ may be introduced into (\ref{eqn: flucceqn2}) and its acoustic component. 
The fluctuation scalar momentum potential gradient and its acoustic components are related through a Poisson equation:
\begin{align}\label{eqn: decomposition3}
(\frac{\partial^2}{\partial x^2}+\frac{\partial^2}{\partial y^2}-\beta^2)\hat{\psi}=-i \omega\hat{\rho}\\
(\frac{\partial^2}{\partial x^2}+\frac{\partial^2}{\partial y^2}-\beta^2)\hat{\psi}_A =-i\omega\frac{\hat{p}}{c^2},
\end{align}
where $\hat{\rho}$ is density  and $\hat{p}$ is the pressure, respectively, of the response mode. 
The Poisson equations for $\hat{\psi}$ and $\hat{\psi}_A$ are solved by prescribing Dirichlet boundary conditions along the cavity wall boundaries. 
For simplicity, the thermal component is calculated using $\hat{\psi}_T=\hat{\psi}-\hat{\psi}_A$. As shown below, this successful decomposition  of response momentum potential into hydrodynamic, acoustic and thermal components provides a much deeper understanding of the flow-acoustic input-output characteristics. 

\section{Results}
\label{sec:results}
\subsection{Stability analysis of mean flow linear operator}
The discounting technique~\citep{jovanovic2004modeling,yeh2018resolvent} used to highlight the short time scale of energy amplification, introduces a free parameter $\kappa$ into the resolvent operator:
\begin{equation}
\mathcal{R}'=[-\text{i}(\omega+\text{i}\kappa) \mathcal{I} - \mathcal{L}({\bm{\bar q}},\beta)]^{-1}\mathcal{M}
\end{equation}
Since $\kappa$ is selected to be slightly larger than the most unstable growth rate~\cite{Yeh:PRF20}, a reasonable choice is obtained from stability analysis, after solving the associated eigenvalue problem.
\begin{figure*}
\centering
\includegraphics [width=0.8\textwidth,trim={0 0.0cm 0 0cm},clip]{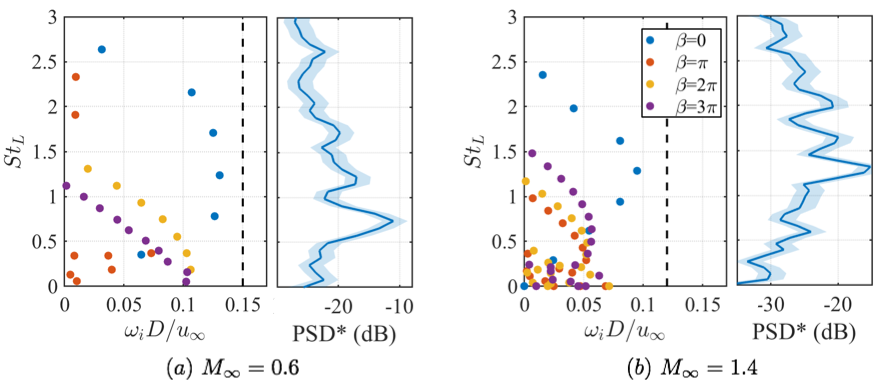}
\caption{Mean flow stability analysis for estimating the discounting parameter $\kappa$ for subsonic and supersonic flows. The dashed line indicates the chosen value of $\kappa$=0.15 and 0.12 for subsonic and supersonic cases, respectively. Power spectral density of instantaneous pressure at $[x, y, z]/D = [3, 0, 0]$ from nonlinear LES simulations.}
\label{fig:eigenspectra}	
\end{figure*}
FIG.~\ref{fig:eigenspectra} shows the eigenspectra of the mean flow stability analysis for both subsonic and supersonic cavity flows. 
The power spectral densities of instantaneous pressure at $[x, y, z]/D = [3, 0, 0]$, as obtained from LES, are shown to the right of each eigenspectrum plot. 
The identified nonlinear oscillations correspond to frequencies obtained from shear layer modes at $\beta=0$. 
This asymptotic analysis shows that the mean flow is unstable for both flow conditions, with 
$\beta=0$ yielding the most unstable modes, except at very low frequencies.

Using this mean flow stability analysis information, we select the same offset parameter $\epsilon=0.025$ for both $M_\infty=0.6$ and $M_\infty=1.4$ cases. 
The discounting parameter is then $\kappa= \max(\lambda_i)+\epsilon$, where $\max(\lambda_i)$ is the most unstable growth rate from the stability analysis. 
Thus, the discounting parameters are $\kappa=0.15$ and $0.12$ for subsonic and supersonic cases, respectively, as indicated by the black dashed lines in FIG.~\ref{fig:eigenspectra}. 
Based on this discounted resolvent operator, we now consider the compressibility effect on the energy amplification of flow systems.

\subsection{Compressibility effect on input-output characteristics}
In the cavity flow, the most important flow-acoustic resonance is linked to the evolution of the shear layer oscillations~\cite{krishnamurty1955acoustic,Rossiter1964,rowley2006dynamics,Beresh:JFM16}. 
The relevant frequencies may be calculated using the modified empirical formula~\cite{Rossiter1964, heller1971flow}, which reads 
\begin{equation}
St_L=\frac{fL}{u_\infty}
 =\frac{n-\alpha}{1/k + M_\infty /\sqrt{1+(\gamma-1)M_\infty^2/2}}
\end{equation}
where $k$ (=0.65) and $\alpha$ (=0.38)~\cite{zhang2019suppression} are the average convective speed of disturbances in shear layer and phase delay, respectively. 
The cavity length $L$ is used to define dimensionless Strouhal number $St_L$, $\gamma=1.4$ is specific heat ratio, and $n=1,2,...$ denotes the $n$th resonant frequency tone of the Rossiter mode. 
The resolvent analysis, and subsequent FT decomposition, performed below examine the input-output characteristics at various frequencies from this spectrum to understand the  corresponding modal features.

Two-dimensional forcing displays the largest growth in the stability analysis of the linearized solution about the mean turbulent state.
The response for the $M=0.6$ case also shows most amplified gain peaks associated with 2D  Rossiter mode frequencies.
However, the most amplified resolvent mode is oblique ($\beta \ne 0$) at $M_\infty=1.4$.
Several experimental campaigns~\citep{zhuang2006supersonic, lusk2012leading, zhang2019suppression} have shown that introducing 3D effects into the cavity flow significantly suppress noise tones. 
As such, the fluid-acoustic response is more amenable to 3D forcing.  
We thus focus the rest of the analysis on $\beta \neq 0$.

\begin{figure*}
\centering
\includegraphics [width=0.8\textwidth,trim={0 0cm 0 0cm},clip]{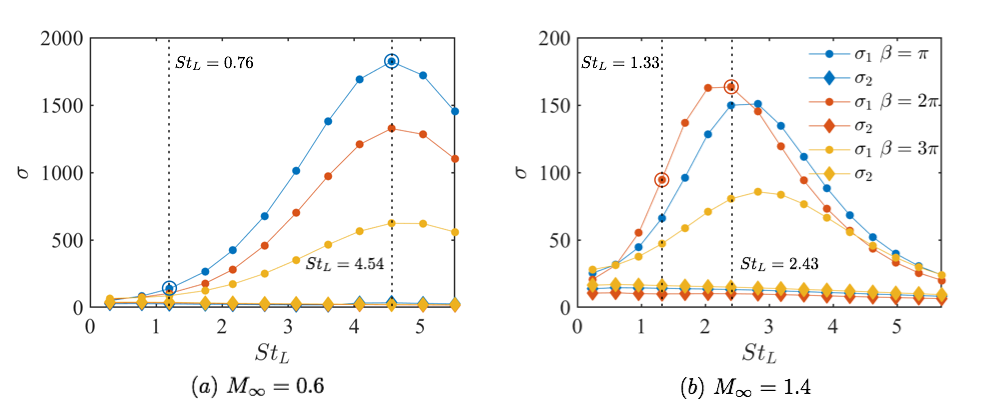}
\caption{Primary ($\sigma_1$) and second ($\sigma_2$) energy amplification at (a) $M_\infty=0.6$ and (b) $M_\infty=1.4$. 
The dominant oscillation detected from LES and higher energy amplification related frequency in resolvent analysis are indicated by dashed lines.}
\label{fig:gain}	
\end{figure*}

The overall energy amplification is a strong function of Mach number.
FIG.~\ref{fig:gain} displays results for the primary and second gain ($\sigma_1$ and $\sigma_2$) for three spanwise wavenumbers, $\beta=\pi$, $2\pi$ and $3\pi$. 
As the Mach number increases from $0.6$ to $1.4$, the overall energy amplification decreases.
The peak energy amplification is an order of magnitude larger for the subsonic flow relative to the supersonic case.
This observation is consistent with several studies of the effect of compressibility in turbulent shear flow. 
\citet{Sarkar:JFM95} showed that the reduced turbulent growth due to compressibility effects is primarily associated with reduced levels of turbulence production. 
In cavity flows also, \citet{Beresh:JFM16} showed a substantial drop in all three components of the turbulence intensity as well as the turbulent shear stress with an increase in Mach number.

In general, both flows are more receptive to high-frequency forcing. 
At $M_\infty=0.6$, the forcing frequency related to the gain peak, $St=4.54$, lies in the vicinity of the $10th$ Rossiter mode frequency. 
This is much higher than the dominant frequency of $St_L= 0.76$ detected in the turbulent cavity LES~\cite{sun2018effects}. 
A similar scenario is observed in the supersonic cavity flow; the frequency associated with the largest energy amplification, $St=2.43$, while lower than the corresponding value for the subsonic case, is also much higher than the strong oscillation frequency tone $St_L=1.33$ detected naturally in the turbulent flow~\cite{liu2020unsteady}. 
The non-normal nature of the operator and non-linear effects thus influence the energy amplification magnitude and its associated forcing frequency in the flow response.

\begin{figure*}
\centering
\includegraphics [width=0.8\textwidth]{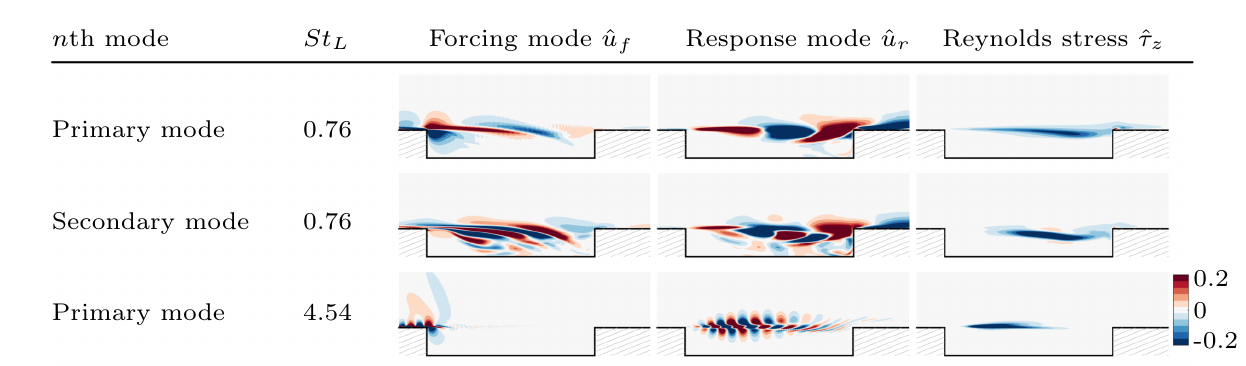}
\caption{Streamwise velocity component of forcing and response modes at $\beta=\pi$ for the cases of $M_\infty=0.6$. The corresponding energy amplifications are indicate as blue circles in figure~\ref{fig:gain}(a).}
\label{fig:ModeM06}	
\end{figure*}

Next, we examine the influence of forcing frequency $St_L$ on the features of the response and forcing modes.
Results for the subsonic case are shown in FIG.~\ref{fig:ModeM06}.
For the sake of conciseness, we elaborate on cases at two frequencies of special interest.  
The first, also designated the natural frequency, $St_L=0.76$, corresponds to the dominant oscillations observed in the pressure fluctuation PSD of the LES; for this, both the primary and secondary modes are shown.  The second frequency, $St_L=4.54$ is the forcing frequency that displays peak gain; only the primary mode is shown here.
At the lower frequency, the dominant regions of both primary and secondary forcing and response modes extend over much of the shear layer.
In contrast, the primary forcing mode at the higher frequency is compactly restricted to the region near the lip.  
The response mode has smaller structures, as anticipated, and are limited to the region around the shear layer.
The region inside the cavity is primarily a region of support for the secondary mode at $St_L=0.76$.
At higher frequency, the secondary modes (not shown) become variants of the primary modes with little presence inside the cavity.
Meanwhile, the gap in the energy amplification significantly enlarges between the primary and secondary modes, as shown in FIG.~\ref{fig:gain}(a). 
The spanwise Reynolds stress of the response mode may be obtained from $\hat{\tau}_z={\mathscr{R}}(\hat{u}_r^{\ast}\hat{v}_r)$, where $^\ast$ is a complex conjugate and $\mathscr{R}$ denotes the real component of the argument. 
Whereas the highest values occur closer to the downstream cavity face, the primary distribution of the Reynolds stress moves upstream in the shear layer, indicating that mixing in the shear layer occurs earlier as frequency increases.

\begin{figure*}
\centering
\includegraphics [width=0.8\textwidth]{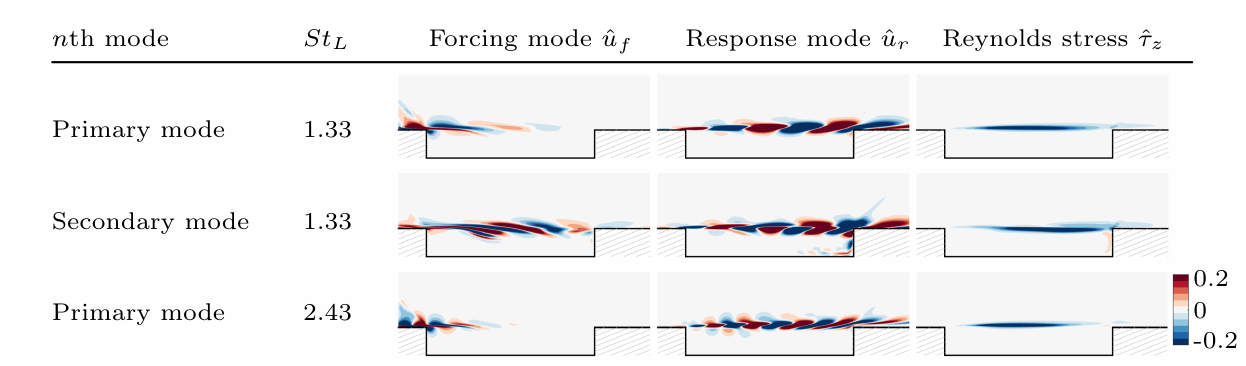}
\caption{Streamwise velocity component of forcing and response modes at $\beta=2\pi$ for the cases of $M_\infty=1.4$. The corresponding energy amplifications are indicate as red circles in figure~\ref{fig:gain}(b).}
\label{fig:ModeM14}	
\end{figure*}

FIG.~\ref{fig:ModeM14} displays forcing and response modes at $St_L=1.33$ and $2.43$ for $M_\infty=1.4$; again, these frequencies correspond to that prominent in the LES, as well as that associated with the resolvent analysis.
However, unlike for $M_\infty=0.6$, the $\beta=2\pi$ results are shown since this wave number shows larger amplification (see FIG.~\ref{fig:gain}).
Some of the features evident in the subsonic flow when frequency is varied are also evident at this supersonic speed in the forcing and response modes.
For instance, the primary response mode is prominent in the shear layer region and its streamwise length scale shortens with increasing forcing frequency. The structures of secondary modes extend inside the cavity. 
At the higher frequencies pertinent to the supersonic case, the structures of response modes extend over the cavity shear layer and do not show much support inside the cavity. Although a similar structure observed for the secondary mode (not shown), its energy amplification is much lower than that of the primary modes, as shown in FIG.~\ref{fig:gain}(b). 
Changing the spanwise wavenumber has a similar effect on the structures of modes across subsonic and supersonic cases; specifically, as the spanwise wavenumber increases, the modal structures become more compact in the wall-normal extent.

\subsection{Fluid-thermodynamic content of resolvent modes}

\begin{figure*}
\centering
\includegraphics [width=0.8\textwidth,trim={0 0cm 0 0cm},clip]{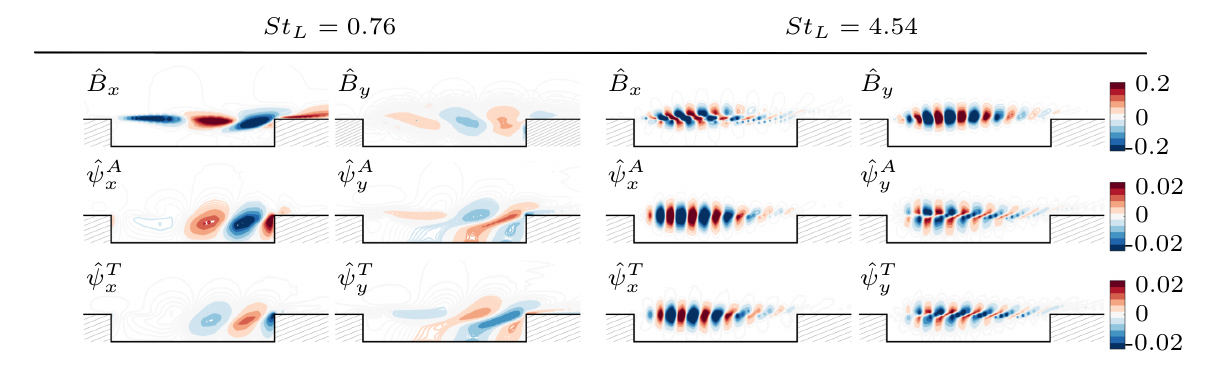}
\caption{Hydrodynamic, acoustic and thermal components of the primary response modes for $\beta=\pi$ at $M_\infty=0.6$.}
\label{fig:M06FTmode}	
\end{figure*}

As mentioned earlier, since the flow-acoustic interaction plays a critical role in self-sustained oscillations, a characterization of the modal fluctuations into  their hydrodynamic,  acoustic and thermal (FT) components is helpful to understand the interaction mechanism and the response of actuator based forcing. 
Doak's decomposition is now employed to characterize the FT content of the primary response mode for the different forcing frequencies.

The three FT components are shown in FIG.~\ref{fig:M06FTmode} for the $M_\infty=0.6$ case at $St_L=0.76$ and $4.54$.  
Since the decomposition is performed on a vector quantity, $\rho \bm{u}$, the FT components are also vectors.  
Therefore, they are plotted separately for the $x$ and $y$ directions; the notation uses subscripts so that, for example, $\hat{\psi}_x^A = \partial \hat{\psi}^A / \partial x$. 
The hydrodynamic components, $\hat{B}$, of the structure of the resolvent response mode much is larger than the acoustic and thermal components. 
This is consistent with observations in other flows, including free jets~\cite{unnikrishnan2016acoustic} as well as hypersonic boundary layer transition~\cite{unnikrishnan2019interactions}.
In fact, at $St_L=0.76$, the streamwise component ($\hat{B}_x$) resembles the primary response mode of streamwise velocity in FIG.~\ref{fig:ModeM06}, consistent with streamwise velocity fluctuations being most prominent in the LES. 
While smaller in magnitude than the hydrodynamic content, the acoustic and thermal components resemble each other, but with a phase shift of $\pi$. 
They are substantially different from any of the forcing or response modes from the resolvent analysis, which have clearly defined structures that extend in the interior of the cavity.
The $y$ components are inclined in the general direction of the mean shear; this is opposite to that observed for $\hat{B}_y$. The inclined pattern of the acoustic mode resembles the radiated acoustic field identified from shorter cavities, such as when $L/D=2$~\citep{Colonius:AIAA01} in which intense structures occur at the trailing edge with an angle of approximately 145 degrees from the streamwise direction.

When the frequency is increased, $St_L=4.54$, the hydrodynamic component exhibits smaller structures over shear layer regions, which are damped near the downstream wall. 
Unlike the results of $St_L=0.76$, where $\hat{B}_x$ dominated, here $\hat{B}_x$ and $\hat{B}_y$ are similar in amplitude.  
Similar conclusions may be made in the acoustic and thermal patterns.  
The acoustic structure arises near the leading edge and progressively decays towards the trailing edge.
This observation is consistent with the sensitivity of the initial shear layer leading to the cavity displaying more sensitivity to higher frequency~\cite{ShawAIAA98}.
The distinct acoustic structures between the lower and higher-frequency cases indicate the different roles of forcing in the flow-acoustic feedback loop. 
A key conclusion thus is that the role of the trailing edge impingement becomes weaker for the high-frequency forcing. 

\begin{figure*}
\centering
\includegraphics [width=0.8\textwidth,trim={0 0cm 0 0cm},clip]{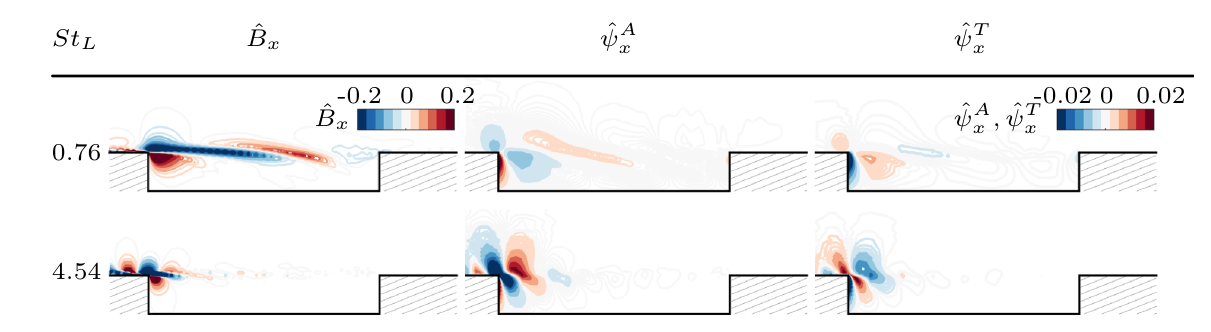}
\caption{Streamwise component of hydrodynamic, acoustic and thermal of primary forcing modes for $\beta=\pi$ at $M_\infty=0.6$.}
\label{fig:M06ForcingFTmode}	
\end{figure*}

The corresponding forcing modes also show the hydrodynamic component dominating the forcing mode at $M_\infty=0.6$. 
 FIG.~\ref{fig:M06ForcingFTmode} shows the FT forcing mode at $\beta=\pi$ and $St_L=0.76$ and $4.54$. The hydrodynamic component pertains to the structure magnitude which is 10 times larger than the other two components.
As the frequency increases, the prominent hydrodynamic structures become smaller and move upstream. 
At $St_L=4.54$, the distribution of hydrodynamic forcing is rather disjoint from the response mode, which suggests the strong convective energy amplification mechanism. In contrast, for acoustic and thermal components, the dominant structures emerge near the leading edge and extend into the cavity, attaching the front wall of the cavity. 
Similar to the response modes, the acoustic and thermal components resemble each other, but with a phase shift of $\pi$.

\begin{figure*}
\centering
\includegraphics [width=0.8\textwidth,trim={0 0cm 0 0cm},clip]{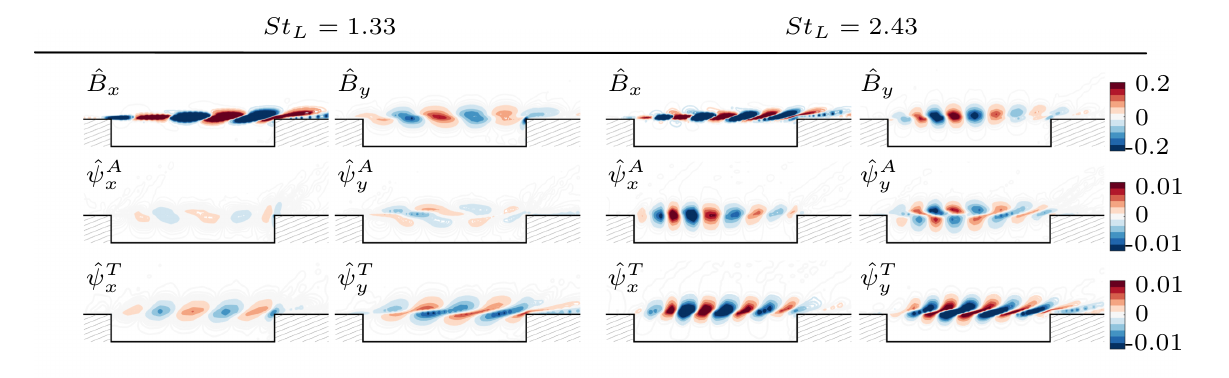}
\caption{Hydrodynamic, acoustic and thermal components of the primary response modes for $\beta=2\pi$ at $M_\infty=1.4$.}
\label{fig:M14FTmode}	
\end{figure*}

Results for  $M_\infty=1.4$ are shown in FIG.~\ref{fig:M14FTmode}. Although the focus here is on $\beta=2\pi$, some qualitative observations remain valid from the $M_\infty=0.6$ case.
At the lower frequency, $\hat{B}_x$ is the most dominant component and is generally contained in the region of the mean shear layer.
Similarly, the number of amplified structures increases with frequency.
Key differences pertinent to control response sensitivity at the two Mach numbers, to be discussed later, also become apparent.
Despite the relatively smaller gap between the natural and resolvent-derived frequencies at $M_\infty=1.4$, the acoustic component of the response shows much greater sensitivity than at $M_\infty=0.6$.
In fact, at the lower frequency, the regular organized structures inside the cavity are not as well defined and the response is relatively weak.
However, the organized structures are very evident at $St_L=2.43$ in wavepacket-like form, which have vertical lobes over much of the cavity, before aligning in the mean shear direction near the trailing edge. 
Similarly, unlike at  $M_\infty=0.6$, where acoustic and thermal features matched each other, at  $M_\infty=1.4$ the thermal structure is quite different from the acoustic mode, as shown.

\begin{figure*}
\centering
\includegraphics [width=0.8\textwidth,trim={0 0cm 0 0cm},clip]{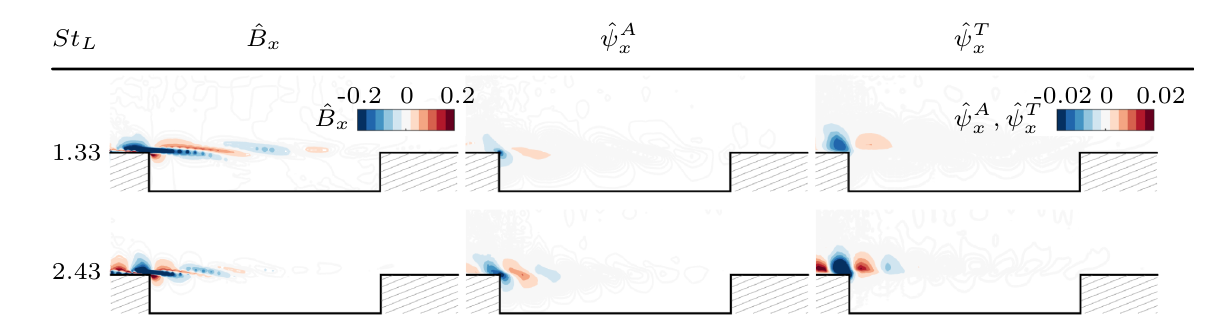}
\caption{Streamwise component of hydrodynamic, acoustic and thermal of primary forcing modes for $\beta=2\pi$ at $M_\infty=1.4$.}
\label{fig:M14ForcingFTmode}	
\end{figure*}

As for the subsonic cases, the length scales of the FT forcing modes are  affected by frequency. 
FIG.~\ref{fig:M14ForcingFTmode} shows the FT forcing modes of $\beta=2\pi$ and $St_L=1.33$ and 2.43 at $M_\infty=1.4$.
The acoustic and thermal components are oriented directly with the shear layer and no substantial structures appear inside the cavity.
The typical length scale reduces but the dominant areas remain unchanged.
This manifests a significantly different behavior than the subsonic flow case $M_\infty=0.6$. 
The isolated forcing modes trigger the response modes around shear layer region which may related to the observation from  \citet{williams2007supersonic}. Supersonic flows may be more likely to operate in the linear regime than subsonic flows since there less overlap inside the cavity and shear layer.
These differences in FT components due to compressibility and frequency now facilitate an analysis of the impact of each localized forcing component, in the context of flow control.

\subsection{Fluid-thermal response to localized-componentwise forcing}
A control strategy may be designed based on the information of the response output and forcing input.
The primary forcing modes show significant structures near the leading edge of the cavity for both subsonic (FIG.~\ref{fig:ModeM06}) and supersonic cases (FIG.~\ref{fig:ModeM14}). 
As such, forcing around the leading edge of the cavity or on the front wall is a natural choice, with the potential to significantly modify the behavior of shear layer, as is the case in most studies to date ~\citep{cattafesta2008active}. 
Inspired by these cavity control studies, we consider localization of the input forcing to two locations: (i) the leading edge $(-0.1\leq x/D \leq 0~\text{and}~0 \leq y/D \leq 0.2)$ and (ii) the front wall $(0 \leq x/D \leq 0.1~\text{and}~-0.2\leq y/D \leq0)$, as indicated in insets of FIG.~\ref{fig:GainM06local}. 

The component-wise forcing input is selected by prescribing constraints in the coupling matrix $\mathcal{M}$ in equation~(\ref{eqn: Resolvent}).
The forcing contains five components expressed as $\hat{\bm f}=[\hat{\rho}_f, \hat{u}_f, \hat{v}_f, \hat{w}_f, \hat{T}_f]$ in equation~(\ref{eqn:decom_NS}), of which 
the spanwise velocity component forcing $\hat{w}_f$ is neglected. 
In the coupling matrix $\mathcal{M}$, the elements at the location of interest are set to the chosen component forcing, which localizes the input term $\mathcal{M} \hat{\bm f}$. 
The matrix $\mathcal{M}$ thus serves two purposes, spatial restriction and imposition of a componentwise forcing filter. 
The results of the resolvent analysis and Doak’s momentum potential theory are discussed in the following to identify the influence of different types of forcing at these two locations on energy amplification and response structures. 
Emphasis is placed on the hydrodynamic and acoustic components which are of interest for cavity flow control. 
  
\subsubsection{Subsonic cavity flow}
\begin{figure*}
\centering
\includegraphics [width=0.8\textwidth]{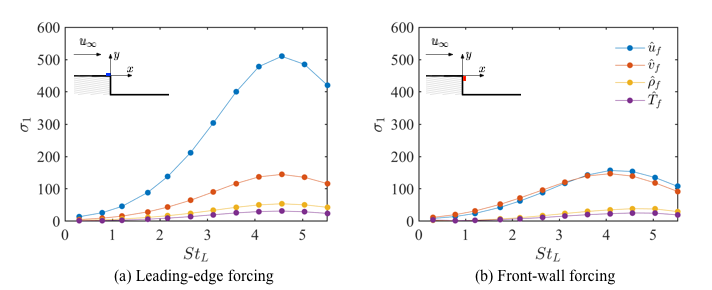}
\caption{Primary gain $\sigma_1$ for (a) leading-edge forcing and (b) front-wall forcing with $\beta=\pi$ at $M_\infty=0.6$.}
\label{fig:GainM06local}	
\end{figure*}
FIG.~\ref{fig:GainM06local} shows energy amplification results for leading-edge and front-wall forcing at $M_\infty=0.6$. 
Considering the former location first, streamwise velocity forcing results in the strongest energy amplification, following by the  $\hat{v}_f$, $\hat{\rho}_f$ and $\hat{T}_f$ forcing in that order. 
The localization of  $\hat{u}_f$ forcing greatly reduces the energy amplification, to about one third that with the global forcing shown in figure~\ref{fig:gain}(a). 
Nonetheless, compared to the global forcing structures in FIG.~\ref{fig:ModeM06}, the much smaller localized forcing at a position where an actuator may actually be placed, attains substantial energy amplification. 
More importantly, the actual frequency value where the peak response is observed, $St_L=4.56$ is the same, suggesting the promising potential from a practical standpoint for substituting the localized forcing for global forcing.

The energy amplification of streamwise velocity forcing $\hat{u}_f$ substantially reduces as the forcing location is changed from the leading edge to the front wall of the cavity, as shown in FIG.~\ref{fig:GainM06local}(b).
This change highlights the importance of leading-edge forcing in the energy amplification mechanism. 
However, one key difference from leading edge forcing is that the streamwise and vertical velocity forcing trigger comparable energy amplification, with the frequencies at which the peaks occur being slightly lower ($St_L=4.08$ as opposed to $4.56$ for the leading edge forcing case).  
The energy amplification for both density and temperature forcing cases remain much lower than the corresponding velocity-wise forcing. 
Overall, these differences reflect the fact that the strongest impact of the forcing location on energy amplification occurs with $\hat{u}_f$ rather than on any of the other variables.

\begin{figure*}
\centering
\includegraphics [width=0.8\textwidth,trim={0 0cm 0 0cm},clip]{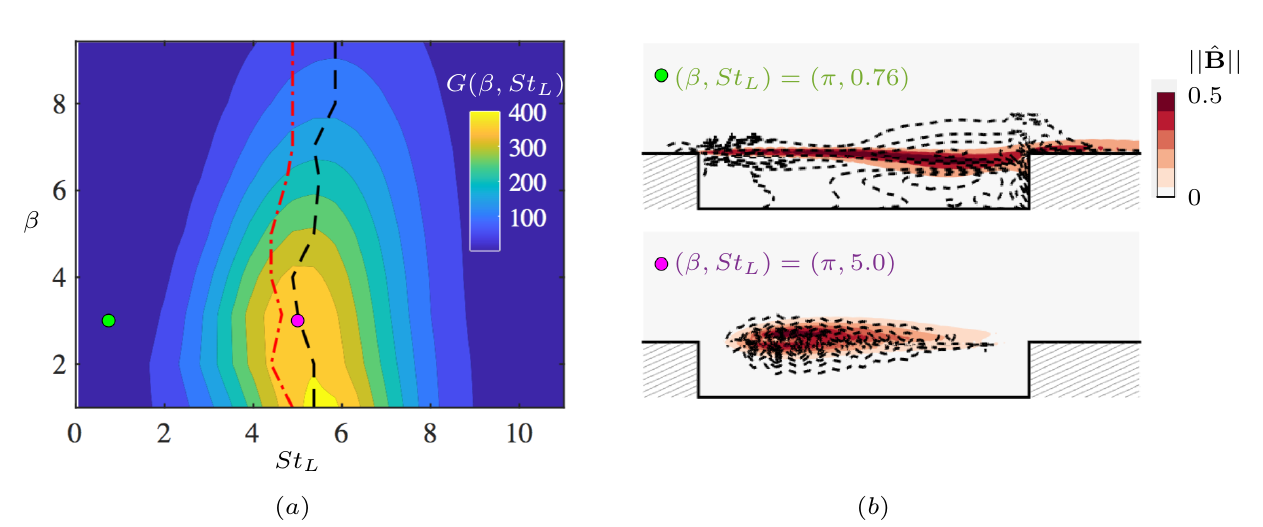}
\caption{(a) Ratio $G(\beta,St_L)$ for leading-edge velocity forcing over frequencies and spanwise wavenumbers at $M_\infty=0.6$. The black dashed line and red dash-dotted line label the peaks of $G(\beta, St_L)$ and $\sigma_1$, respectively. (b) Spatial distribution of hydrodynamic $||\hat{B}||$ (contour plot) and acoustic $||\hat{\psi}^A_{\bm x}||$ (black dashed lines) structures at $(\beta,St_L)=(\pi,0.76)$ (top) and $(\beta,St_L)=(\pi,5.0)$ (bottom). The locations of their $G(\beta, St_L)$ are indicated in green and magenta dots on plot (a).}
\label{fig:M06LEmode}	
\end{figure*}

The relative sensitivity of the acoustic versus hydrodynamic response may be assessed through the ratio
\begin{equation}
G(\beta,St_L)=\frac{\sigma_1(\beta, St_L)||\hat{\psi}^A_{\bm x}(\beta, St_L)||}{||\hat{\bm B}(\beta, St_L)||},
\end{equation}
where $||\cdot||$ denotes the magnitude of FT components and $\sigma_1(\beta, St_L)$ is the primary energy amplification. The magnitude $||\hat{\psi}^A_{\bm x}||$ combines the acoustic components in $x$ and $y$ direction.
As a representative sample, the quantification of $G(\beta,St_L)$ over spanwise wavenumber $\beta$ and frequency, for the case of leading edge imposition of $\hat{u}_f$ perturbations is shown in FIG.~\ref{fig:M06LEmode}(a). 
The black line traces the frequency where the ratio $G(\beta,St_L)$ is maximum for each spanwise wave number, while the red line indicates the frequency where the overall energy amplification is highest. 
The peak corresponding to the largest modification of the acoustic component $G(\beta,St_L)$ occurs at $(\pi, 5.0)$, which is higher than that associated with the energy amplification peak  ($St_L=4.54$). 
As the spanwise wavenumber increases, this peak associated forcing frequency remains relatively unchanged (black dashed line in FIG.~\ref{fig:M06LEmode}(a)).

Further details of the response  may be obtained by examining the spatial distributions of the FT modes.
FIG.~\ref{fig:M06LEmode}(b) shows $||\hat{B}||$ and $||\hat{\psi}^A_{\bm x}||$ at two frequencies.
The top figure is that associated with the natural frequency $St_L=0.76$ (marked with a green dot in frame (a)) while the bottom figure corresponds to the frequency $St_L=5.0$, related to the $G(\beta,St_L)$ peak (magenta dot).
In each, $||\hat{B}||$ are represented by the flooded contours, while $||\hat{\psi}^A_{\bm x}||$ are displayed with dashed contours.
For $St_L=0.76$, the hydrodynamic features span approximately the region of the shear layer from the leading edge to the trailing edge.
However, acoustic structures also show a presence within the cavity, as well as on the side of the flow near the trailing edge.
The peak associated with $G(\beta, St_L)$ is connected to overlapping acoustic and hydrodynamic structures.  
At this higher frequency, the strongest hydrodynamic and acoustic structures move forward closer to the leading edge. 
This overlap is a feature of the location where the peak is observed in $G(\beta, St_L)$ \textit{i.e.,}  overlapping acoustic and hydrodynamic response components result in larger modifications to the acoustic components. 
The exchange of energy between hydrodynamic and acoustic components underlying this result follow a complex dynamics that may be described by the dynamics of the total fluctuating enthalpy~\cite{jenvey1989sound, prasad2021nature}.  
The analysis is beyond the scope of the current effort; results in the context of jets have been presented by \citet{unnikrishnan2019interactions}.

\begin{figure*}
\centering
\includegraphics [width=0.8\textwidth,trim={0 0cm 0 0cm},clip]{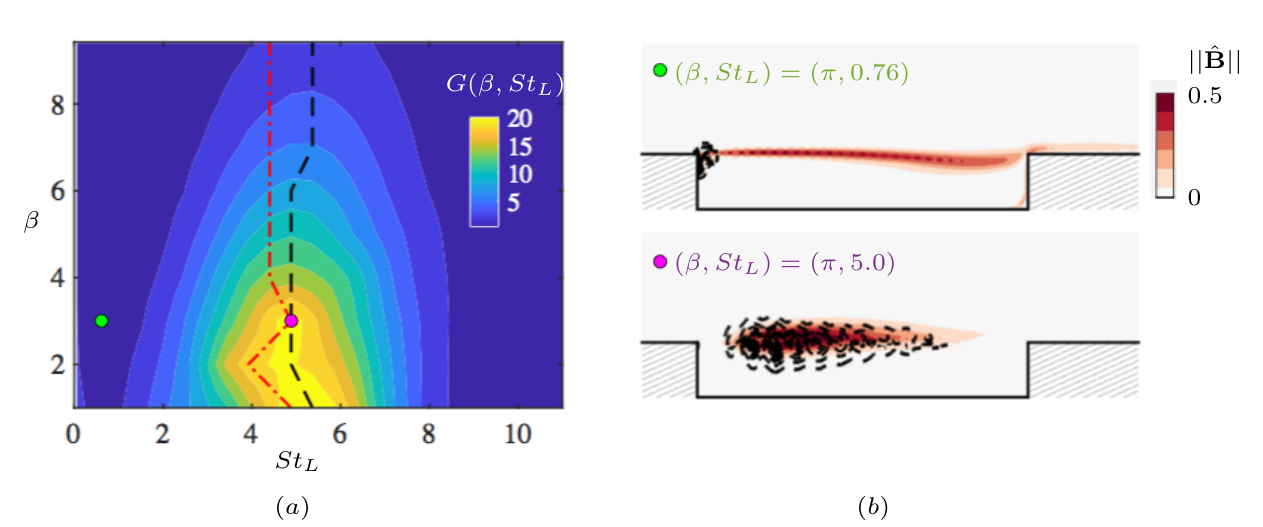}
\caption{(a) Ratio $G(\beta,St_L)$ for front-wall temperature forcing at different frequencies and spanwise wavenumbers for $M_\infty=0.6$. The black dashed line represents the peak related frequency over spanwise wavenumbers and red dash-dotted line labels the peak of $\sigma_1$ related frequencies over spanwise wavenumbers. (b) Spatial distribution of hydrodynamic $||\hat{B}||$ (contour plot) and acoustic $||\hat{\psi}^A_{\bm x}||$ (black dashed lines) structures at $(\beta,St_L)=(\pi,0.76)$ (top) and $(\beta,St_L)=(\pi,5.0)$ (bottom). The locations of their $G(\beta, St_L)$ are indicated in green and magenta dots on plot (a).}
\label{fig:M06FWTmode}	
\end{figure*}     

To illustrate the results with another type of forcing, we consider the case of $\hat{T}_f$ excitation on the front-wall, which displays different local acoustic structures.
The $G(\beta, St_L)$ contours and corresponding structures of $||\hat{B}||$ and $||\hat{\psi}^A_{\bm x}||$ are shown in FIG.~\ref{fig:M06FWTmode}. 
The $G(\beta, St_L)$ peak related frequency is generally higher than the frequency related to the $\sigma_1$ peak, which suggests a strong effect on the acoustic response may not align with the energy amplification at the two frequencies of interest. 
Examining FIG.~\ref{fig:M06FWTmode}(b), at $St_L=0.76$, the local intense acoustic response occurs around the forcing location, while, the hydrodynamic response structure remains over the shear layer region. 
When the forcing frequency increases, the peak related response structure in FIG.~\ref{fig:M06FWTmode}(b) resembles the case of $\hat{u}_f$ shown in FIG.~\ref{fig:M06LEmode} at $St_L=5.0$, which has an overlapping area over the shear layer. 
It is noteworthy that except for the difference in the value of $G(\beta, St_L)$, that at the same high-frequency, the different forcing types have similar acoustic and hydrodynamic response structures. 
This observation holds for all cases examined, and indicates that the high-frequency selection mechanism is less influenced by the location or type of forcing.

\subsubsection{Supersonic cavity flow}

\begin{figure*}
\centering
\includegraphics [width=0.8\textwidth]{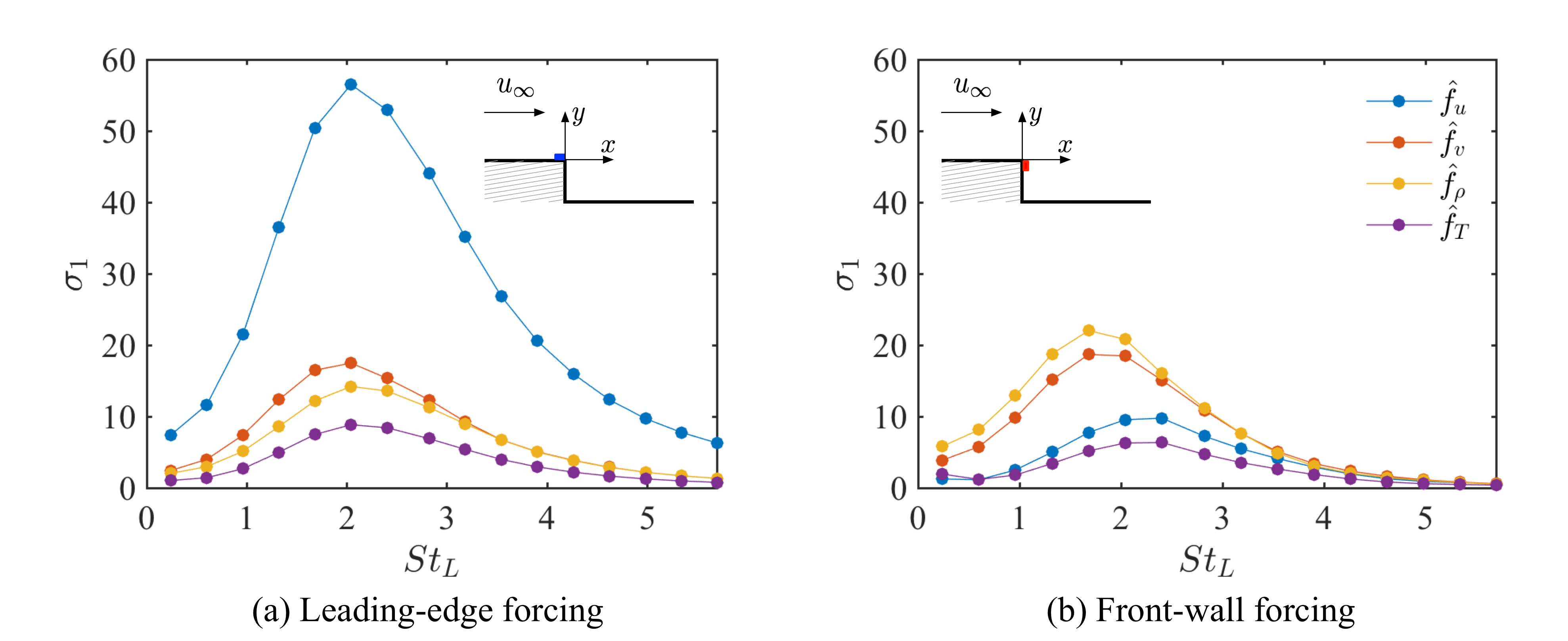}
\caption{Primary gain $\sigma_1$ for (a) leading-edge forcing and (b) front-wall forcing with $\beta=2\pi$ at $M_\infty=1.4$.}
\label{fig:GainM14local}	
\end{figure*}

\begin{figure*}
\centering
\includegraphics [width=0.8\textwidth,trim={0 0cm 0 0cm},clip]{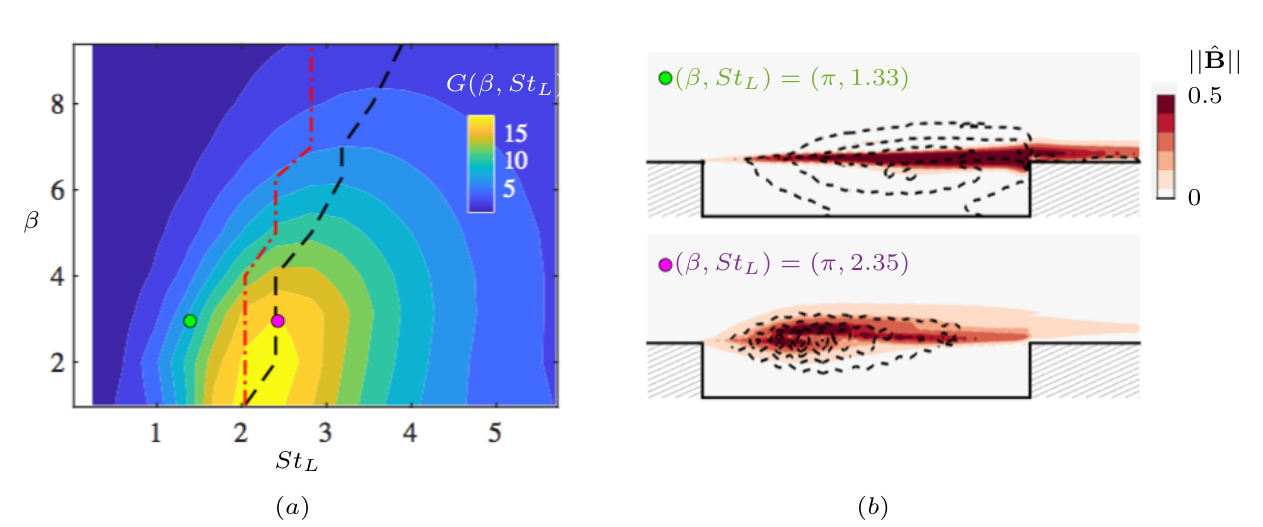}
\caption{(a) Ratio $G(\beta,St_L)$ for leading-edge velocity forcing at different frequencies and spanwise wavenumbers in the case of $M_\infty=1.4$. The black dashed line and and red dash-dotted line represent the peaks $G(\beta,St_L)$ and $\sigma_1$ related frequency over spanwise wavenumbers, respectively. (b) Spatial distribution of hydrodynamic $||\hat{B}||$ (contour plot) and acoustic $||\hat{\psi}^A_{\bm x}||$ (black dashed lines) structures at $(\beta,St_L)=(\pi,1.33)$ (top) and $(\beta,St_L)=(\pi,2.35)$ (bottom). The locations of their $G(\beta, St_L)$ are indicated in green and magenta dots on plot (a).}
\label{fig:M14FWmodeGain}	
\end{figure*}

As noted for the subsonic case, localized forcing in the supersonic case also reduces the energy amplification relative to the original unconstrained forcing.
However, many features, including its distribution over frequency and spanwise wavenumber resemble that of the global case (shown earlier in FIG.~\ref{fig:gain}(b)). FIG.~\ref{fig:GainM14local} shows the energy amplification for leading-edge and front wall forcing cases for $M_\infty=1.4$.  
As for $M_\infty=0.6$, changing forcing location from the leading-edge to the front-wall again reduces the energy amplification due to $\hat{u}_f$. 
In contrast, the effect of the other types of forcing is negligible. 
Based on these similarities and differences, we discuss the distribution of $G(\beta,St_L)$ and response structures of leading-edge $\hat{u}_f$ and front-wall $\hat{T}_f$ in detail. 

The distribution of $G(\beta,St_L)$ indicates stronger acoustic response at lower spanwise wavenumbers. 
FIG.~\ref{fig:M14FWmodeGain}(a) shows the variation of $G(\beta, St_L)$ with frequency and spanwise wavenumber. 
As the spanwise wavenumber increases, the value of $G(\beta,St_L)$ decreases, and the peak related frequency gradually shifts to higher frequencies, as indicated by the black dashed line in FIG.~\ref{fig:M14FWmodeGain}. 
The corresponding structures related to the natural frequency $St_L=1.33$ and $St_L=2.35$ for $\beta=\pi$ are shown in FIG.~\ref{fig:M14FWmodeGain}(b) with the magnitudes of $||\hat{B}||$ and $||\hat{\psi}^A_{\bm x}||$ as before. 
At $St_L=1.33$, the strong acoustic component occurs at the rear part of the cavity with a vertical extent expanding into the cavity. 
The corresponding hydrodynamic structure gradually spreads over the shear layer. 
For the case of $St_L=2.35$, both structures are more diffuse, but there is significant overlap in the shear layer region. 
This feature resembles the observation for subsonic cases where $G(\beta, St_L)$ peaks at higher frequencies.

\begin{figure*}
\centering
\includegraphics [width=0.8\textwidth,trim={0 0cm 0 0cm},clip]{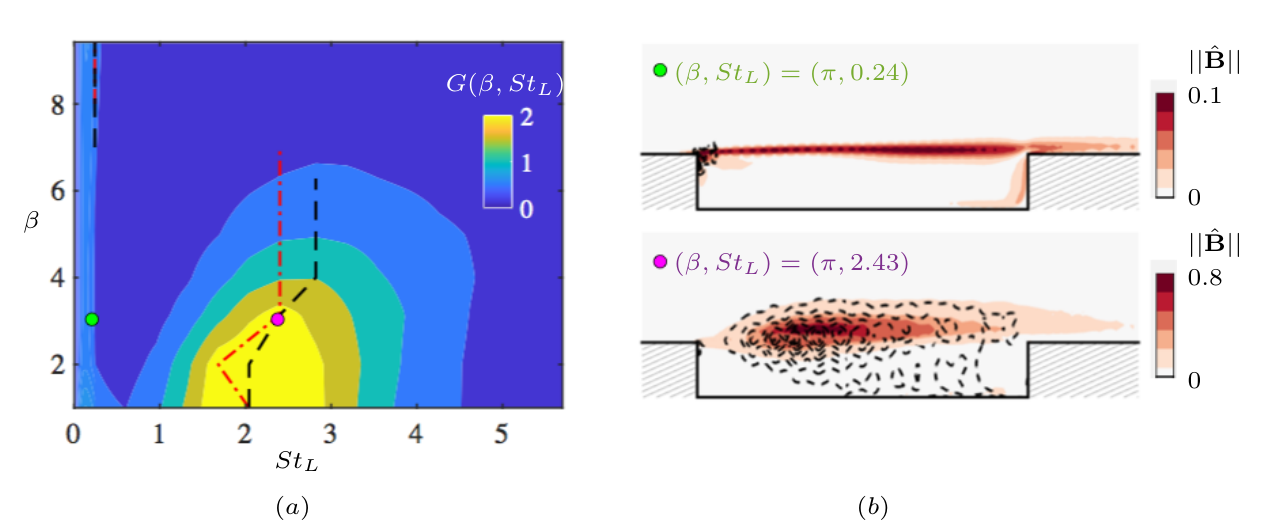}
\caption{(a) Ratio of $G(\beta,St_L)$ for front-wall temperature forcing over frequencies and spanwise wavenumbers at $M_\infty=1.4$. (b) The hydrodynamic $||\hat{B}||$ (contour plot) and acoustic $||\hat{\psi}^A_{\bm x}||$ (black dashed lines) components for two representative cases at $(\beta, St_L)=(\pi, 0.24)$ and $(\pi, 2.43)$. The black dashed line and red dash-dotted line in (a) represent the peak $G(\beta,St_L)$ and $\sigma_1$ related frequency over spanwise wavenumbers, respectively. The locations for two representative cases at $(\beta, St_L)=(\pi, 0.24)$ and $(\pi, 2.43)$ are indicated with green and magenta dots, respectively.}
\label{fig:M14FWT}	
\end{figure*}

Besides a relatively lower magnitude of $G(\beta, St_L)$ for the case of front-wall $\hat{T}_f$, its corresponding acoustic structures are distinct from other cases.
FIG.~\ref{fig:M14FWT} shows the distribution of $G(\beta, St_L)$, magnitude of $||\hat{B}||$ and $||\hat{\psi}^A_{\bm x}||$ at $St_L=0.24$ and $2.43$, which are representative cases to elaborate the unique acoustic response to front-wall temperature forcing.
Overall, the values of $G(\beta, St_L)$ are much smaller than for $\hat{u}$ forcing.
The slightly higher value of  $G(\beta, St_L)$ emerges for the case with low frequency $St_L<0.5$. A representative acoustic and hydrodynamic structure pair for $St_L=0.24$ is shown in FIG.~\ref{fig:M14FWT}(b) (upper figure). 
The localized intense acoustic structures appear where forcing is executed. However, the hydrodynamic component remains spread out over the cavity due to the convective nature. 
For structures related to the $G(\beta, St_L)$ peak  (lower figure), the dominant acoustic component occurs inside the cavity and overlaps with the hydrodynamic component.
These  features are common with acoustic and hydrodynamic distributions associated with $G$ peak at different forcing cases. 
The overall low magnitude of $G$ across frequency and spanwise wavenumber suggests the front-wall $\hat{T}_f$ forcing is ineffective at triggering hydrodynamic and acoustic response. 
   
\section{Concluding Remarks}
\label{sec:conclusion}
We have combined resolvent analysis and Doak's momentum potential theory to investigate the input-output characteristics of subsonic ($M_\infty=0.6$) and supersonic flows ($M_\infty=1.4$) past an $L/D=6$ cavity at $Re=10,000$. 
The goal is to identify input-output characteristics of the system in terms of their specific acoustic and hydrodynamic components, with a view towards a deeper understanding of the flow-acoustic type of flow problems.
The time- and spanwise-averaged flow derived from an LES, prescribed spanwise wavenumbers and frequencies are used as inputs for the resolvent analysis. 
The resolvent analysis confirms that the energy amplification in the supersonic cavity flow is much lower than that from the subsonic case, which is consistent with the effect of increasing compressibility effect in the turbulent shear flow.
Moreover, the frequency associated with the peak energy amplification is different from the natural value \textit{i.e.,} that prominent in the corresponding LES, for both subsonic and supersonic cases. 
Regardless of Mach number, the acoustic component appears at the trailing edge for the case with natural frequency. 
As the forcing frequency is increased, the intense regions of the acoustic mode moves upstream and decays progressively towards the trailing edge. 
Thus, high-frequency forcing where the resolvent analysis predicts peak amplification induces different acoustic responses compared to natural frequency forcing.

The coupling matrix is used to isolate the influence on energy amplification and acoustic response of the different forcing types, such as streamwise, wall-normal velocity, thermal or density perturbations, and actuator localization to accessible regions, specifically the leading edge or the front wall.
For both Mach numbers, leading-edge forcing achieves more extensive energy amplification than the front-wall case. 
The ratio between the acoustic and hydrodynamic component $G(\beta,St_L)$ indicates strong acoustic modification. 
The related peak frequency is higher than that associated with maximum energy amplification. 
At spanwise wavenumbers and frequencies where $G(\beta,St_L)$ is maximum, the acoustic and hydrodynamic response structures overlap in the region of the shear layer. 
Overall, the current effort provides a deeper understanding of the flow-acoustic input-output characteristics from a linear perspective. 
Future investigation will focus on the identification and differentiation of linear observations in the nonlinear LES cavity flows.  

\section*{Acknowledgments}
 The authors gratefully acknowledge the support of AFOSR (FA9550-19-1-0081; Program Officer: G. Abate) and ONR (N00014-17-1-2584, Program Officer: S. Martens). The authors would like to acknowledge Dr. K. Taira from University of California, Los Angeles and Dr. Y. Sun from Syracuse University for providing time-averaged cavity flow data.

\end{document}